# The information content of Local Field Potentials: experiments and models

Alberto Mazzoni [a], Nikos K. Logothetis [b, c], Stefano Panzeri [a,d] *

[a] Center for Neuroscience and Cognitive Systems, Istituto Italiano di Tecnologia, Via Bettini 31, 38068 Rovereto (TN), Italy

[b] Max Planck Institute for Biological Cybernetics, Spemannstrasse 38, 72076 Tübingen, Germany

[c] Division of Imaging Science and Biomedical Engineering, University of Manchester, Manchester M13 9PT, United Kingdom

[d] Institute of Neuroscience and Psychology, University of Glasgow, Glasgow G12 8QB, United Kingdom

* Corresponding author: S. Panzeri, email: stefano.panzeri@glasgow.ac.uk

# 1. Local Field Potential dynamics offer insights into the function of neural circuits

The Local Field Potential (LFP) is a massed neural signal obtained by low pass-filtering (usually with a cutoff low-pass frequency in the range of 100-300 Hz) of the extracellular electrical potential recorded with intracranial electrodes. LFPs have been neglected for a few decades because in-vivo neurophysiological research focused mostly on isolating action potentials from individual neurons, but the last decade has witnessed a renewed interested in the use of LFPs for studying cortical function, with a large amount of recent empirical and theoretical neurophysiological studies using LFPs to investigate the dynamics and the function of neural circuits under different conditions.

There are many reasons why the use of LFP signals has become popular over the last 10 years. Perhaps the most important reason is that LFPs and their different band-limited components (known e.g. as alpha, beta or gamma bands) are invaluable for understanding cortical function. They offer unique windows onto integrative excitatory and inhibitory synaptic processes at level of neural population activity. The LFP captures a multitude of neural processes, such as synchronized synaptic potentials (Mitzdorf, 1985; Logothetis, 2003), afterpotentials of somatodendritic spikes (Gustafsonn, 1984), and voltage-gated membrane oscillations (Harada and Takahashi, 1983; Kamondi et al., 1998). As a result, the LFP is sensitive to subthreshold integrative processes and carries information about the state of the cortical network and the local intracortical processing, including the activity of excitatory and inhibitory interneurons and the effect of neuromodulatory pathways (Mitzdorf, 1985; Logothetis, 2003, 2008). These contributions are almost impossible to capture using spiking activity from only a few neurons (Logothetis et al., 2002; Buszaki, 2006; Mazzoni et al., 2008). Therefore, the combined recording and analysis of LFPs and spikes offers insights into the circuit mechanism generating neural representation of information that cannot be obtained at present by examining spikes alone.

Another source of interest is that cortical LFPs typically contain a very broad spectrum of oscillations or of fluctuations of neural activity, that span a wide range of frequencies ranging from less than one Hz to one hundred Hz or more (Kayser and Konig, 2004; Lakatos et al., 2005; Buszaki, 2006; Senkowski et al., 2007). This broad band range of activities most likely reflects contribution of several different neural processing pathways. A bulk of literature has shown that these contributions can, to some extent, be separated out and identified from LFPs with relatively simple techniques such as frequency decompositions (Steriade et al., 1993; Logothetis et al., 2002; Buszaki, 2006; Mazzoni et al., 2008; Ray and Maunsell, 2010) Therefore, recording LFPs allows the empirical examination and separation of different and potentially independent information channels participating in neural information processing (Belitski et al,2008;Montemurro et al,2008; Kayser et al,2009).

An additional reason of interest of LFPs is that they provide stable signal for a longer period of time than multiunit spiking activity, and are therefore useful for long-term chronic experiments and for clinical applications such as Brain Machine Interfaces (Mehring et al., 2003; Andersen et al., 2004; Rickert et al., 2005; Scherberger et al., 2005; Donoghue, 2008; Nicolelis and Lebedev, 2009; Bansal et al., 2011; Markowitz et al., 2011).

However, LFPs have also a drawback. Due to the multiple neuronal processes that contribute to them, the LFP is a partly ambiguous signal and is more difficult to interpret than spikes. To take full advantage of the opportunities this signal offers to study neural information processing and cortical organization, careful analytical, modeling and empirical considerations are therefore necessary. To illustrate the challenges to a computational understanding of LFPs, and to present the progress we made so far, here we review a series of neurophysiological (Belitski et al., 2008; Montemurro et al., 2008) and modelling (Mazzoni et al., 2008; Mazzoni et al., 2010) studies from our group which attempted to identify and separate out, respectively through extracellular recordings from the primary visual cortex of anaesthetized macaques and through simulations of recurrent networks, the neural pathways involved into the primary cortical representation of naturalistic visual information and the type of information carried by each pathway. In particular, and taking inspiration from a recent review of our group about these topics (Mazzoni et al., 2011),, we aim to show and discuss how models and experiments contribute to give a coherent understanding of how LFP recordings may be used to study neural population coding.

## 2. Information content of primary cortical Local Field Potential during naturalistic visual stimulations

As we mentioned above, the LFPs is a broadband signal that captures variations of neural population activity over a wide range of time scales (Kayser and Konig, 2004; Lakatos et al., 2005; Buszaki, 2006; Senkowski et al., 2007). The range of time scales available in LFPs is particularly interesting from the neural coding point of view, because, it opens up the possibility to investigate whether there are privileged time scales for information processing, a question that has been hotly debated over the last one or two decades. On the one hand, the presence of a wide spectrum of activity could imply that there is no privileged scale for information representation, because information is evenly spread over all scales (Panzeri et al., 2010). This view is consistent with the proposal that neural activity is largely scale free (Bedard et al., 2006; Mazzoni et al., 2007; He et al., 2010). On the other hand, it is possible that information is represented by only a small number of specific frequency ranges, each carrying a separate contribution to the information representation. To shed light on this issue, it is important to quantify the information content of each frequency range of neural activity, and understand which ranges carry complementary or similar information.

To investigate this question, we used spectral analysis and information theory to analyse LFPs and spiking responses recorded from the primary visual cortex of anaesthetized macaques in response to binocularly-presented naturalistic colour movies. Here we used naturalistic movies for several reasons. First, they reflect better, in a better way than simplified stimuli do, the range of continuous changes over a wide range of spatial and temporal scales that characterize the natural visual environment (Reinagel, 2001; Simoncelli, 2003; Felsen and Dan, 2005; Geisler, 2008). And perhaps more importantly for the present study, the use of complex stimuli containing many parameters varying partly independently (as it happens in naturalistic movies) gives us a better chance to engage different processing pathways in a partly independent way, and this facilitates the separation, through analysis and models, of different processing channels from LFP recordings.

In the following, we first review the experimental methods used to collect the data and we then present the results of a number of computational analyses aimed at characterizing quantitatively the information carried by different frequency components of the spectrum of extracellular recordings.

## 2.1 Neurophysiological procedures and spectral analysis of extracellular signals

We briefly summarize the experimental procedures used to record neural responses to naturalistic colour movies in primary visual cortex (V1). We refer to previous (Belitski et al., 2008; Montemurro et al., 2008; Belitski et al., 2010; Mazzoni et al., 2011) for full details.

Four adult rhesus monkeys (Macaca mulatta) participated in the experiments. All procedures were approved by the local authorities (Regierungsprsidium) and were in full compliance with the guidelines of the European Community (EUVD 86/609/EEC) for the care and use of laboratory animals. Prior to the experiments, form-fitted head posts and recording chambers were implanted during an aseptic and sterile surgical procedure (Logothetis et al., 2002). Recordings were obtained while the animals were anaesthetized. The main reason for collecting neural responses during anaesthesia is that this protocol offers several advantages for the investigation of primary cortical dynamics. In particular: anaesthesia removes effects of attention and arousal, as well as trial to trial variations of eye movements, and good signal-to-noise ratio, which can be more easily obtained in anesthetized animals due to long acquisition times. Neuronal activity was recorded from opercular V1 (foval and para-foveal representations) using microelectrodes (FHC Inc., Bowdoinham, Maine, 300–800k Ohms) which were arranged in a 4 x 4 square matrix (interelectrode spacing varied from 1 mm to 2.5 mm) and introduced in each experimental session into the cortex through the overlying dura mater by a microdrive array system (Thomas Recording, Giessen, Germany). Electrode tips were typically positioned in the upper or middle cortical layers. In each of the 5 recording sessions we recorded from 7-10 sites in V1 with a well-defined receptive field within the field of movie projection for a total of 45 sites.

Visual stimuli were presented binocularly at a resolution of 640×480 pixels (field of view: 30×23 degrees, 24 bit true color, 60 Hz refresh) using a fiberoptic system (Avotec, Silent Vision, Florida). Stimuli consisted of commercially available Hollywood movies (30 Hz frame rate) displaying 'naturalistic' image dynamics, from which 3.5–6 min long sequences were presented and repeated 30–40 times. The receptive fields of all recording sites analysed were within the area of visual stimulation (Rasch et al., 2008).

Extracellular neural signals were recorded at a 20.83 KHz rate. LFPs were extracted by low-pass-filtering the neural signal in the frequency range up to 250Hz. In order to extract multiunit spike times, the neural signal was bandpassed in the high-frequency range of 500-3500 Hz. The threshold for spike detection was set at 3.5 standard deviations. A spike was recognized as such only if the last spike occurred more than one millisecond earlier. For the present analysis we did not separate single and multi-units. Power spectra were computed in single trials using the multitaper technique (Percival and Walden, 1993) which provides an efficient way to simultaneously control the bias and variance of spectral estimation.

### 2.2 Spectral properties of LFP recordings in response to natural stimuli

To quantify the characteristics of LFP fluctuations at different frequencies, we first computed the LFP spectrum during the entire period in which movies were presented (an example channel, reporting LFP response spectra to two different movies and LFP spectra in absence of visual stimulation is shown in Fig 1A). The LFP spectrum had a very wide band with fluctuations ranging over the entire frequency range analysed. The highest LFP power was at low frequencies (< 10 Hz), and the power decreased steeply at increasing frequencies. We compared the averaged LFP spectrum evoked during the movie to the LFP spectrum of the same electrode during spontaneous activity (measured in the absence of visual stimulation). There was a proportionally small increase of power during movie stimulation at frequencies below 10 Hz (Fig. 1A). The evoked and spontaneous LFP spectrograms were similar at frequencies between 10and 24 Hz, while the power associated to frequencies higher than approximately 40 Hz was higher during movie presentation. The difference in power between different movies was also more marked in the region [40-100 Hz]. Consistent with previous studies (Henrie and Shapley, 2005), we found the most substantial power increase over spontaneous activity of the movie-evoked LFP in the gamma frequency region [40-100 Hz].

We next analyzed the reliability of the responses of different LFP band to the movie presentation. Figs. 1B-D show the bandpassed LFP responses (in the [1-5 Hz], [28-32 Hz] and [72-76 Hz] frequency range respectively) for several trials, from a representative example recording site. The presentation of the movie elicited LFP patterns that, both in the low frequency (1-5Hz) range (Fig. 3B) and in the [72-76 Hz] range within the high gamma region (Fig. 1D), were clearly modulated by the movie and repeatable across trials: episodes of high instantaneous power were elicited reliably in correspondence of certain scenes in the movie. In contrast, LFP waveforms in the intermediate frequency range [28-32 Hz] (Fig. 1C) could not be reliably associated to the movie time course.

We also investigated the V1 spiking responses to the movie. Figure 1E shows that the spike rates clearly encoded the movie time course. The high spike rate episodes were associated more closely with episodes of high LFP power in the high-gamma LFP frequency range than at lower LFP frequencies, suggesting that gamma LFPs may be more closely related to the stimulus-modulated spiking activity than low LFP frequencies.

### 2.3 Information analysis of LFPs suggest that cortex multiplexes naturalistic information over a small number of frequency bands

We then investigated more quantitatively how the power of different LFP frequency bands encoded the stimuli computing the information that the LFP power carries about which scene was being presented. Mutual information (abbreviated as "information" in the following), is a popular measure of the goodness of stimulus encoding in neuroscience (de Ruyter van Steveninck et al., 1997; Fairhall et al., 2001; Panzeri et al., 2003). It quantifies the reduction of the uncertainty about the stimulus that can be gained from observing a single-trial neural response, and we measured it in units of bits (1 bit means a reduction of uncertainty by a factor of two). We refer to Chapter 9 for a more in-depth discussion about mutual information and its application to Neuroscience.

Information depends on both the choice of the stimulus set and of the quantification of the neural response. To create the stimulus set S, we divided the presentation time of the dynamic stimulus (a movie in the case of recordings from visual cortex, and a time dependent input spike train in the case of neural network simulations) into different non-overlapping segments of length T (a parameter that was varied in the range from few ms to several seconds (Belitski et al., 2008)) and each segment was considered as a different stimulus s (see schematic in Fig. 2). We then computed the information about which dynamic stimulus time segment elicited the considered response. This procedure has several advantages. The first is that it is simple to apply and lends itself to comparisons between experimental and theoretical data. The second is that it does not make any assumption as to which specific features of the dynamic stimulus triggered the neural response and so can potentially capture the information about all possible dynamical stimulus features presented experimentally (de Ruyter van Steveninck et al., 1997). Regarding the choice of neural response R, we considered several different possibilities, as detailed next.

We first computed the information between the stimulus and the power of the LFP at a given frequency f, as follows:

$$I(S;R_f) = \sum_s P(s) \sum_{r_f} P(r_f|s) \log_2 \frac{P(r_f|s)}{P(r_f)} \tag{0.1}$$

where *P(s)* is the probability of presentation of the stimulus window *s* (here equal to the inverse of the total number of stimulus windows ), *P(r_f)* is probability of observing power $r_f$ across all trials in response to any stimulus (Fig. 2A) and *P(r_f |s)* is the probability of observing a power $r_f$ at frequency *f* in response to a single trial to stimulus *s* (Fig. 2C-D). To facilitate the sampling of response probabilities, the space of power values at each frequency was binned into 6 equipopulated bins (Belitski et al., 2008). For all measures of information about power presented here, we used a stimulus window of length T = 2.048 s.

The information of the LFP power, averaged over all channels recorded in session A98 (the one selected for comparison with model results, see below), is reported in Fig 3A. We found two informative bands in the LFP spectrum: a low frequency range below 10 Hz (corresponding to the delta and theta bands) and the [40 – 100 Hz] in the gamma band.

Having established that both high-gamma and low frequencies of LFPs convey information, the next important question is to establish whether the different informative frequencies ranges are redundant or not, *i.e.* whether or not they carry the same or different stimulus information. The above single-frequency information analysis was extended to compute how much information about the stimuli we can obtain when combining together the power $r_{f1}$ and $r_{f2}$ at two different frequencies. The mutual information that the joint knowledge of the powers $r_{f1}$ and $r_{f2}$ conveys about the stimulus is defined as:

$$I(S;R_{f1}R_{f2}) = \sum_s P(s) \sum_{r_{f1}r_{f2}} P(r_{f1}r_{f2}|s) \log_2 \frac{P(r_{f1}r_{f2}|s)}{P(r_{f1}r_{f2})} \tag{0.2}$$

where the stimulus-response probabilities defined in Eq (1.2.) are analogous to the ones in Eq (1.1.) – see (Belitski et al., 2008) for full details. The redundancy between the information carried by the powers at the two considered frequencies is defined as the difference between the sum of the information carried by the power of each frequency individually and their joint information:

$$Red(S;R_{f1}R_{f2}) = I(S;R_{f1}) + I(S;R_{f2}) - I(S;R_{f1}R_{f2}) \tag{0.3}$$

Thus, we computed both the joint information carried by the power of pairs of LFP frequencies (Eq. 1.2) and their redundancy (Eq. 1.3). We found that the redundancy between the information carried by the power of high and low frequencies was nearly zero (Fig. 3B). Consequently, the information obtained by the combined knowledge of the power at low frequencies and the power at gamma frequencies was nearly the sum of the information carried by the two frequencies separately (Fig. 3C). In contrast, frequencies in the gamma band were highly redundant between each other (Fig. 3C), suggesting that all frequencies in the gamma range reflect largely the same network phenomenon. We found also that low LFP frequencies carried independent information with respect to spike rates, and were indeed totally decoupled from spike rates (both in terms of stimulus selectivity and trial to trial covariations). However, the power of gamma range LFPs was largely (but not completely) redundant to the spike rate, suggesting that spike rates and gamma power are a largely overlapping information channel (results not reported here but fully explained in (Belitski et al., 2008)).

In summary, V1 LFP spectral information about naturalistic stimuli is multiplexed in two independent streams, one at very low frequencies and one at gamma frequencies. The gamma power carries information partly (but not completely) redundant to that carried by spike rates. Powers at different frequencies in the low frequencies range carry largely independent information, while frequencies in the gamma range are highly redundant with each other. Intermediate frequencies do not carry any information, suggesting that out of the many time scales of neural activity that have significant power, only a handful of time scales seem to carry different types of information.

## 2.4 Coding of information by phase of firing

The previous sections showed that during stimulation with naturalistic dynamics visual cortex develops slow fluctuations which are informative about the external world and can be measured by recording LFP. An interesting question is whether these fluctuations provide an internal temporal frame which can be used to reference spike times and increase the information that can be extracted from them. The existence of this type of spike timing coding, called phase of firing, has received considerable attention in recent years. Evidence has been reported that spatial-navigation- and memory-related structures encode some information by phase of firing (O'Keefe and Recce, 1993; Huxter et al., 2003; Lee et al., 2005). However, the extent to which firing rate and phase encode genuinely different information is still debated (Harris et al., 2002; Mehta et al., 2002; Harris, 2005). Furthermore, the research has focused so far mainly on the role of phase of firing in the hippocampus, without clarifying whether phase coding represents a fundamental code for cortical information transmission used already in primary sensory areas.

In a recent study, we used the V1 recordings of LFPs and spikes in response to movie stimuli to investigate whether such phase of firing codes carry information about complex naturalistic visual (Montemurro et al., 2008). We found that the presentation of naturalistic color movies elicited reliable responses across trials both for the spikes and for the delta-band [1–4 Hz] LFP phase (Fig. 4A top rows). To visualize how LFP phases were modulated by the movie, we divided the phase range into four equi-spaced quadrants and labeled each with a different color (Fig. 4A). It was apparent that the [1–4 Hz] LFP phase also encoded the movie, because the phase values were modulated by the movie time, and this modulation was reliable across trials at several times during the movie (Montemurro et al., 2008). Visual inspection of the data suggested the LFP phase at which spike were fired allowed to disambiguate different movie scenes eliciting the same firing rate (Fig. 4A bottom rows), suggesting that the phase of firing carried visual information not available in the spike rates.

This point was investigated in detail using information theory. We defined a phase of firing code, which is the neural response defined when the timing of spikes emitted in response to the stimulus are measured with respect to the phase of a concurrent LFP wave bandpassed around a considered frequency $f$. This can be done by dividing the phase range into quarters, and then by tagging the spikes with a label indicating the phase quadrant at which they were emitted. Then, the *phase of firing information* can be defined as follows:

$$I(S; R_f^\Phi) = \sum_s P(s) \sum_{\varphi_f=0}^{4} P(\varphi_f | s) \log_2 \frac{P(\varphi_f | s)}{P(\varphi_f)} \tag{0.4}$$

where in the above equation the response set is composed by 5 different symbols: $\varphi_f = 0$ denotes the absence of spikes in response to the stimulus in the considered trial, and $\varphi_f = 1, 2, 3, 4$ denote that a spike was emitted when the LFP phase was in a given quadrant. To evaluate if the phase of firing carries information above and beyond that carried by spike rates, we compared $I(S; R_f^\Phi)$ to the *spike rate information*, which was evaluated from Eq (1.4) but after randomly shuffling the responses $\varphi_f > 0$ independently for each trial. The shuffling operation preserves all the information carried by the spike rate while at the same time completely destroying any additional information that may be carried by the knowledge of phase. For all phase of firing information measures reported here, we used a stimulus window T of 4 ms (Montemurro et al., 2008).

We found (Fig 4B) that the phase of the low frequency [1-4 Hz] LFP at which spikes were fired carried 55% more information than spike counts about the movie segment being shown. Labeling the spikes with the phase of higher frequencies LFPs increased the information by a much smaller amount, suggesting that spike times are particularly informative with respect to slow (rather than fast) LFP fluctuations. In another study, the phase of firing with respect to [4-8 Hz] LFPs was found to carry large amount of information about complex natural sounds in auditory cortex (Kayser et al., 2009), suggesting a general role of such code in representing sensory stimuli with complex, naturalistic dynamics. Slow fluctuations in the excitability of the local network can be measured by considering low frequency LFPs (Logothetis, 2002; Buzsaki and Draghun, 2004; Schroeder and Lakatos, 2009), and the phase of such LFPs reflects the timing of changes in the state of the network and in its excitability. Thus, the increased information available in the phase of firing with respect to low frequency LFP fluctuations suggests that knowledge of the network state generating a spiking response would increase the information that the spiking response carries.

## 3. Multiplexing of information in an integrate and fire model of a local cortical network

The experimental studies described above suggest that visual information might be multiplexed in two separate information streams associated to separate frequency band of the LFP. However these studies do not clarify what are the mechanisms generating this two information channels and the nature of the sensory information encoded in the two bands. In order to investigate which mechanisms are compatible with our observations, we used biophysically plausible neural network models, which we review in the following.

### 3.1. Description of the recurrent models of networks of excitatory and inhibitory neurons

We used a simulated model of a recurrent sparsely connected neural network of leaky integrate and fire (LIF) neurons (Tuckwell, 1988; Brunel and Wang, 2003). These networks are simple, but able to capture the interplay between excitation and inhibition, which is one feature of the organization of cortical microcircuits which is believed to shape the dynamics of local mass activation (Douglas and Martin, 1991; Deco et al., 2008; Logothetis, 2008). The network we used (Fig. 5A) was composed of 4000 *pyramidal* neurons with AMPA-like synapses, and 1000 *interneurons* with GABA-like synapses. AMPA and GABA postsynaptic currents were determined by the spikes emitted by the pre-synaptic neurons of the network and by the external inputs mimicking the thalamic inputs (conveying the information about sensory stimuli) and the ongoing cortical fluctuations (summarizing the effect of the slow covariations in network state due to ongoing activity). Synapses carrying both types of external inputs were activated by random Poisson spike trains, with time-varying rates which were identical for all neurons. The network connectivity was random and sparse, with a connection probability of 0.2 between any pair of cells. We refer to the original reports (Mazzoni et al., 2008; Mazzoni et al., 2010), as well as to Chapter 26, for more details about this work in particular and about models of neural networks in general.

In order to compare simulation and experimental results, we needed to compute simulated LFPs from the model network. The simplest model of LFP consists in averaging the membrane potential of the network neurons (Hill and Tononi, 2005; Xing et al., 2009). This approximation, however, has the disadvantage that it may slow down the synaptic potentials by an extra factor due to the low pass filtering properties of the neural membrane, and that it might be unable in some circumstances to pick effects potentially based on the interplay between inhibitory and excitatory stimuli (Mazzoni et al., 2008; Linden et al., 2010). Since a prominent contribution to real cortical LFPs arises from current flows due to synaptic activity (Mitzdorf, 1985; Logothetis, 2003), a better way to model of LFPs is to use weighted sums of the currents crossing the membrane, whether given only by synaptic components (Esser et al., 2007), or by a wider range of sources (Makarov et al., 2010). In the work presented in this chapter (see (Mazzoni et al., 2008; Mazzoni et al., 2010)) we computed the simulated LFP signal generated by the network as the sum of the absolute values of AMPA and GABA currents (the model does not include other currents). We decided to sum the absolute values of currents because AMPA synapses are usually apical and GABA synapses are usually peri-somatic and thus their dipoles sum with the same sign along the dendrite (Fig. 5B). Our model can then be considered as a two compartments model with a single current on each compartment. Our simple model takes into account a very general feature of the geometry of cortical organization: since pyramidal neurons are likely to contribute more due to their approximate open field arrangement (Murakami and Okada, 2006), in computing the simulated LFPs we summed only currents from synapses of pyramidal neurons (Fig. 5B). The LFP signal was taken with a negative sign to be better compared with the polarity of our experimental recordings. As shown below, we found that the simple quantification that we chose was enough to reproduce all the features of interest of the neural dynamics observed during real primary cortical recordings, thereby suggesting that our simple description of LFPs was sufficient for the present purposes. Interestingly, a bulk of more comprehensive modelling work considering in more detail the geometry of neural organization and how different sources of LFPs combine in the extracellular medium (Einevoll et al., 2007; Pettersen et al., 2008; Linden et al., 2011), concludes that the type of LFP models used here are the simplest way to capture most relevant properties of field potentials (Linden et al., 2010).

### 3.2. Simulation results

Three types of inputs signal were injected, in different simulation sessions, from the model "thalamic" region: (i) time-invariant ("constant") inputs, (ii) perfectly periodic inputs which varied sinusoidally in time, (iii) "naturalistic" input spike trains which reproduced the firing activity recorded in the LGN of an anaesthetized monkey during one of the binocular naturalistic visual stimulation sessions described above (Rasch et al., 2008).

In a first series of simulations in (Mazzoni et al., 2008) we injected the network with constant inputs of different intensity. In agreement with previous studies (Brunel and Wang, 2003), we found that the gamma power of the LFP increased monotonically with input strength (Fig. 6A). These results are consistent with neurophysiological findings that grating stimuli of increasing contrast (which is known to modulate the thalamic input to V1 (Shapley et al., 1981; Derrington and Lennie, 1984)) indeed modulate also the power of the LFP gamma band in V1 (Friedman-Hill et al., 2000; Henrie and Shapley, 2005). Similar relationships between stimuli intensity and gamma band power have also been found in non-invasive recordings in humans (Swettenham et al., 2009; Muthukumaraswamy et al., 2010). Furthermore, in the network model (Mazzoni et al., 2008) increases in the inputs intensity were not only associated to a stronger power of the gamma band, but also to an higher peak frequency, consistent with the experimental recordings from visual cortices (Ray and Maunsell, 2010).

In a second set of simulations we studied the response of the network to time-varying inputs. We injected periodic inputs of different frequency and intensity. The strength by which a given frequency component of the LFP of the simulated network entrained to a given frequency component of the time course of the network spike rate input was quantified by the circular variance (Luo and Poeppel, 2007; Lakatos et al., 2008; Chandrasekaran et al., 2010) over time of the difference between the considered frequency component of the LFP and the considered frequency component of the input (each bandpassed in 2 Hz wide bands with the Kaiser filter detailed above). We found that a sufficiently strong and slow periodic input was able to entrain the network LFP at the corresponding frequency (Fig. 6B). This is compatible with the entrainment between low frequency stimuli and neural activity in the primary visual cortex recently observed in auditory cortices (Fisher, 1993). The level of entrainment was weaker for higher frequencies (Mazzoni et al., 2008). Note that since input peak phases correspond to higher input intensities, they are associated to high gamma power. Therefore the model can account also fo the widely observed phenomenon of cross-frequency coupling (Mazzoni et al., 2010).

In a third and final set of simulations we injected the network with a naturalistic input based on the firing activity recorded from the LGN of an anaesthetized macaque presented with naturalistic movies. The LGN recordings were performed during the same session of the V1 LFP recordings whose spectral information was presented in Fig. 3. The naturalistic LGN spiking response had a spectrum with a strong power associated to a broad range of low frequencies (Mazzoni et al., 2008). Fig. 7 illustrates that the rules found above for simple inputs held also for the broadband naturalistic input: the low frequency LFP entrained to the large-amplitude slow fluctuations of the input, and the power of gamma oscillations increased in correspondence of high input rates. As a consequence, the information carried by the LFP power of the simulated network during naturalistic stimulation closely matched the experimental one (Fig. 8A, compare to Fig. 3A): significant information was conveyed at low frequencies, and in the gamma band and above, with the two information peaks separated by an interval of non-informative frequencies. Also the redundancy structure was similar to the one found experimentally: frequencies above 40 Hz were redundant with each other and independent from the informative frequencies at the low end of the spectrum (Fig. 8B, compare to Fig. 3B). This was a consequence of the fact that input features encoded in the two bands, the intensity and the power at low frequencies, varied independently from scene to scene (Mazzoni et al., 2008). When injecting the network with naturalistic inputs we analysed also the dynamics of the phase of firing code in the network responses. Each input was injected 100 times with different outcomes of the process representing the cortical spontaneous activity in order to study the reliability of the LFP and the amount of information conveyed by the phase of firing code. We bandpassed the network LFP response into different frequency bands and we found that the reliability of the LFP phase was high for low frequencies and reached at ~30 Hz a minimum level that was stable for higher frequencies. The same dynamics was observed in the experimental data (Montemurro et al., 2008), even if frequencies above 30 Hz were more reliable in simulations than in recordings. We built the phase of firing code considering the cumulative spiking activity of small groups of neurons (from 1 to 10), since this is the order of magnitude of the units available from single electrode recordings as those used in (Montemurro et al., 2008). We found that the gain in information of the phase of firing relatively to the spike rate grew linearly with the average firing rate, i.e. the amount of information that the phase label added to each spike was relatively stable.

As observed in (Montemurro et al., 2008), the phase of firing gain was larger for low frequencies (Fig. 8C)) even if it never reached zero in simulations because of the higher reliability of high frequency bands. Notably, very good quantitative agreement with experimental data was obtained when considering the cumulative activity of few excitatory units with a total average firing rate of 5-10 spikes/sec (Fig. 8C), in the range of typical values of rates recorded from a single extracellular electrode.

In sum, the model reproduced well the experimental data and suggested answers to the two questions about the underlying mechanism asked at the beginning of the section: i) the low frequencies conveys information about the low frequency components in the input, and the gamma frequencies convey information about the strength of the input; ii) encoding at low frequencies occurs through entrainment of the local neural activity to the external stimuli, encoding at gamma frequencies occurs through modulations of locally generated rhythms. Note that the presence of two streams of information could only be detected using naturalistic inputs which display large variations in intensity and a power spectrum peaking at low frequencies. Moreover, the same framework and particularly the same network can be used to analyze the phase of firing code described in (Montemurro et al., 2008; Kayser et al., 2009) .

## 4. Discussion

LFPs offer the opportunity to get important insights into the function of cortical microcircuits. However, a full understanding of what they can (or cannot) tell us about the function of microcircuits requires understanding how LFPs originate from cortical neural networks, and how they can be analyzed and modeled. Though this task is obviously a difficult challenge, an important advantage is that of LF (unlike for other measured of neural activity such as fMRI) a reliable biophysical modeling scheme linking activity of individual neurons and measured LFP signals has been established (Holt and Koch, 1999; Einevoll et al., 2007; Mazzoni et al., 2008; Pettersen et al., 2008; Leski et al., 2010; Linden et al., 2010; Mazzoni et al., 2010; Leski et al., 2011; Linden et al., 2011). In this article we presented the progress that we made so far in using these computational models, as well as specialized analytical tools, in characterizing the frequency ranges expressed by primary cortical neural populations, and captured by the LFPs they generate, to represent naturalistic sensory information.

We presented results showing that, out of the whole broad LFP spectrum, only two frequency ranges carry significant information: low frequencies (< 10 Hz) and gamma range. Moreover, the information carried by the two frequency bands is largely independent. Our models suggest that the origin of this information independence stems from the difference in network mechanisms originating the activity in the different frequencies and in the different nature of the information encoded. In particular, our models suggest that inhibitory–excitatory neural interactions generate gamma-range oscillations encoding the strength of the input to the network, while the slow fluctuations of the naturalistic stimuli are encode slow dynamic features of the input into slow LFP fluctuations because the two fluctuations are entrained.

Taken together, these findings suggest that a key strategy used by cortical networks to cope with the challenges of representing the complexity of the natural environment is to use a multiplexing strategy to encode simultaneously different types of information at different time scales and so enhance the information capacity of cortical columns (Panzeri et al., 2010). Evidence is now accumulating that a multiplexing strategy, suggested in earlier seminal work (Bullock, 1997; Lisman, 2005) is key for the brain to represent the complexity of changing environments (Fairhall et al., 2001; Lundstrom and Fairhall, 2006) and of the information relevant for behavior (Schyns et al., 2011). It is interesting to note that we found that the multiplexed information was concentrated over a small number of frequency ranges rather than being uniformly distributed over the entire frequency spectrum. An advantage of encoding information into a limited number of frequency ranges is that it gives the benefit of multiplexing without complicating too much the decoding procedure, which needs to pay attention only to a small number of different frequencies rather than to a continuum of time scales.

Our empirical evidence in support of multiplexing has also potential implications for the development of Brain Machine Interfaces (Donoghue, 2008; Nicolelis and Lebedev, 2009), because decoding based on multiple time scales may be used to enhance the amount of information that can be extracted by each single electrode.

The hypothesis that the information about naturalistic movies carried by low frequency LFPs results from entrainment to the low frequency features of naturalistic stimuli, is coherent with results from the auditory system, showing that low frequency LFPs entrain to sound features during the presentation of complex naturalistic sounds (Chandrasekaran et al., 2010).If this hypothesis is correct low frequency LFPs in primary visual cortex should entrain to the frame by frame changes of one or more image features (such as contrast or orientation) displayed in the receptive field of the considered recording sites during movie presentation. This is because the latter should somehow reflect the time course of the input to the local cortical network, as these features influence the firing of geniculate neurons and of other cortical areas.

Natural visual stimuli are characterized by both "what" aspects (image properties such as contrast or orientation which are defined by the relationship between visual signals simultaneously presented at different points in space) and "when" aspects, describing the temporal variations of the various image features. Although temporal information is crucial for sensory processing, our understanding of stimulus processing is currently biased toward the representation of non-temporal aspects (Buonomano and Maass, 2009). A potentially interesting implication of the hypothesis that low frequency LFPs entrain to the slow stimulus dynamics is that they may provide a way to encode information about the temporal structure of natural visual stimuli in the low temporal frequency range (< 10 Hz) in which these stimuli have the most power and information (Dan et al., 1996; Geisler, 2008).

## Acknowledgements

This work was supported by the Italian Institute of Technology, the Max Planck Society, the Compagnia di San Paolo, and by the SI-CODE project of the Future and Emerging Technologies (FET) programme within the Seventh Framework Programme for Research of The European Commission, under FET-Open grant number:FP7-284553.

# Figures

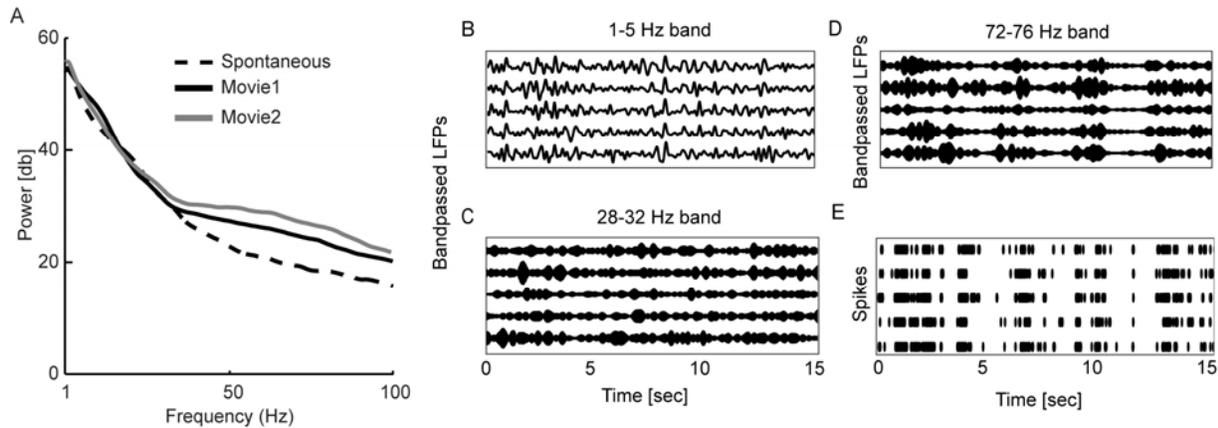

**Figure 1:** *Examples of LFP responses in macaque V1*. (A) Trial-averaged LFP spectrum for a representative channel during presentation of two different movies and during spontaneous activity. (B) LFP traces (band-passed in the [1-5 Hz] frequency range) from five presentations of a 15 s long movie extract. Traces were displaced on the vertical axis to make them distinguishable. (C) Time courses of the [28-32Hz] band-passed LFP recorded during the same five movie presentations of (B). (D) Time courses of the [72-76Hz] band-passed LFP recorded during the same five movie presentations of (B). Frequency bands were obtained bandpassing LFPs with a Kaiser filter (see text). Panels (B)-(D) adapted from (Panzeri et al., 2008). (E) Raster plot of spike times (indicated by black markers) recorded from the same electrode of LFP displayed in (B)-(D) during the same five presentations shown in (B).

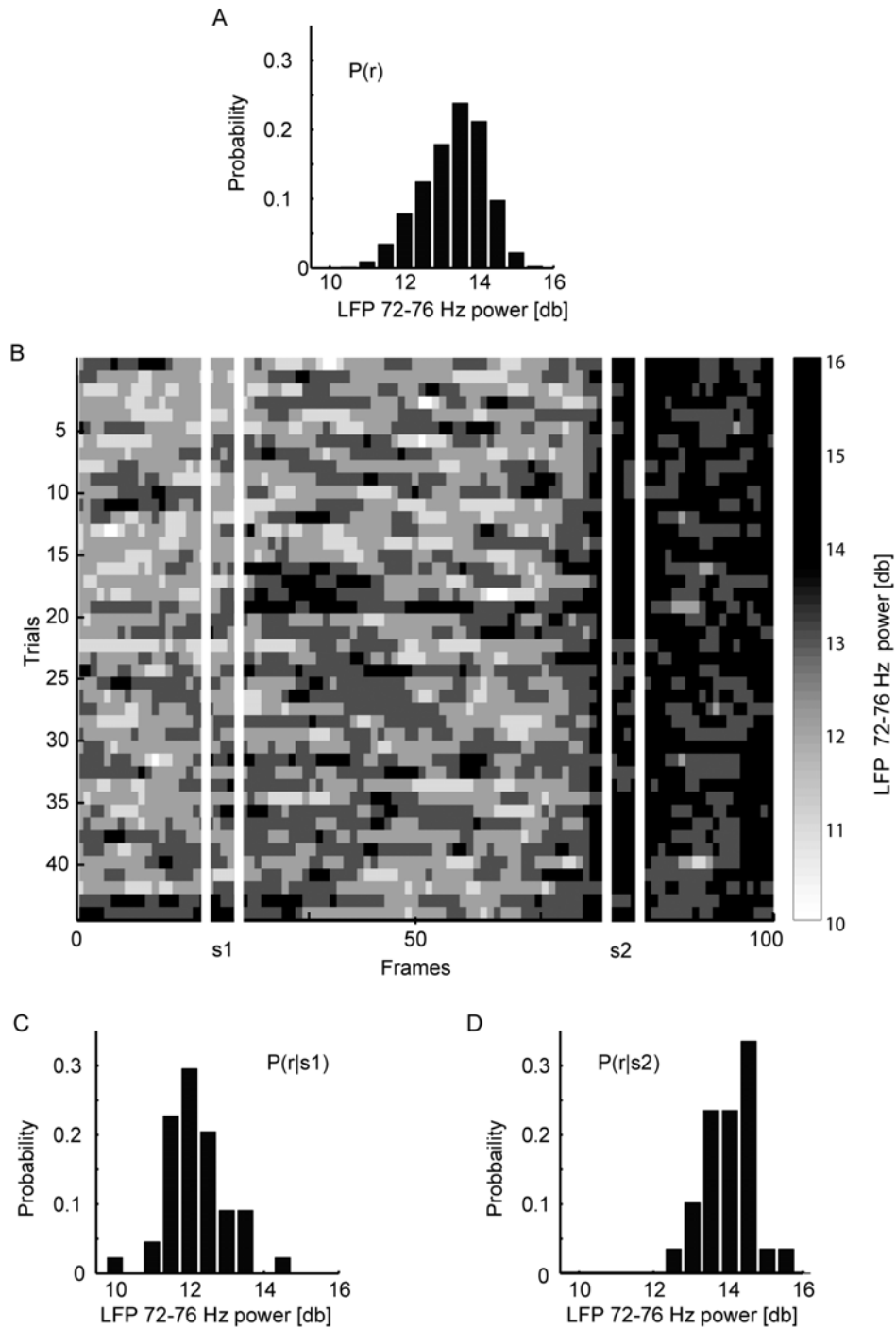

**Figure 2:** *Schematic representation of the computation of the mutual information carried by LFP power about which movie scene was being presented*. The figure illustrates how we obtained the different probabilities needed to compute (through Eq (1.1)) the information $I(S;R)$ about the movie carried by the LFP power at a given frequency. The movie presentation time is portioned into non-overlapping windows, each considered a different stimulus *s* (a "scene"). The set of the stimuli is the set of the N different scenes, each of which is presented once every trial, so the probability of each scene is $P(s) = 1/N$. (A) Probability distribution $P(r)$ of the LFP 72-76 Hz power across all trials and scenes. (B) Single-trial LFP power across all trials and movie scenes. (C) and (D) Probability distribution $P(r|s)$ of the LFP [72-76 Hz] power across trials at the time of presentation of two particular scenes s1 and s2, respectively. The differences between the two distributions and the distribution $P(r)$ suggest that the LFP gamma power carried information about which scene is presented. The key for evaluating mutual information expressed in Eq. (1.1) is the computation of $P(r|s)$ for all scenes.

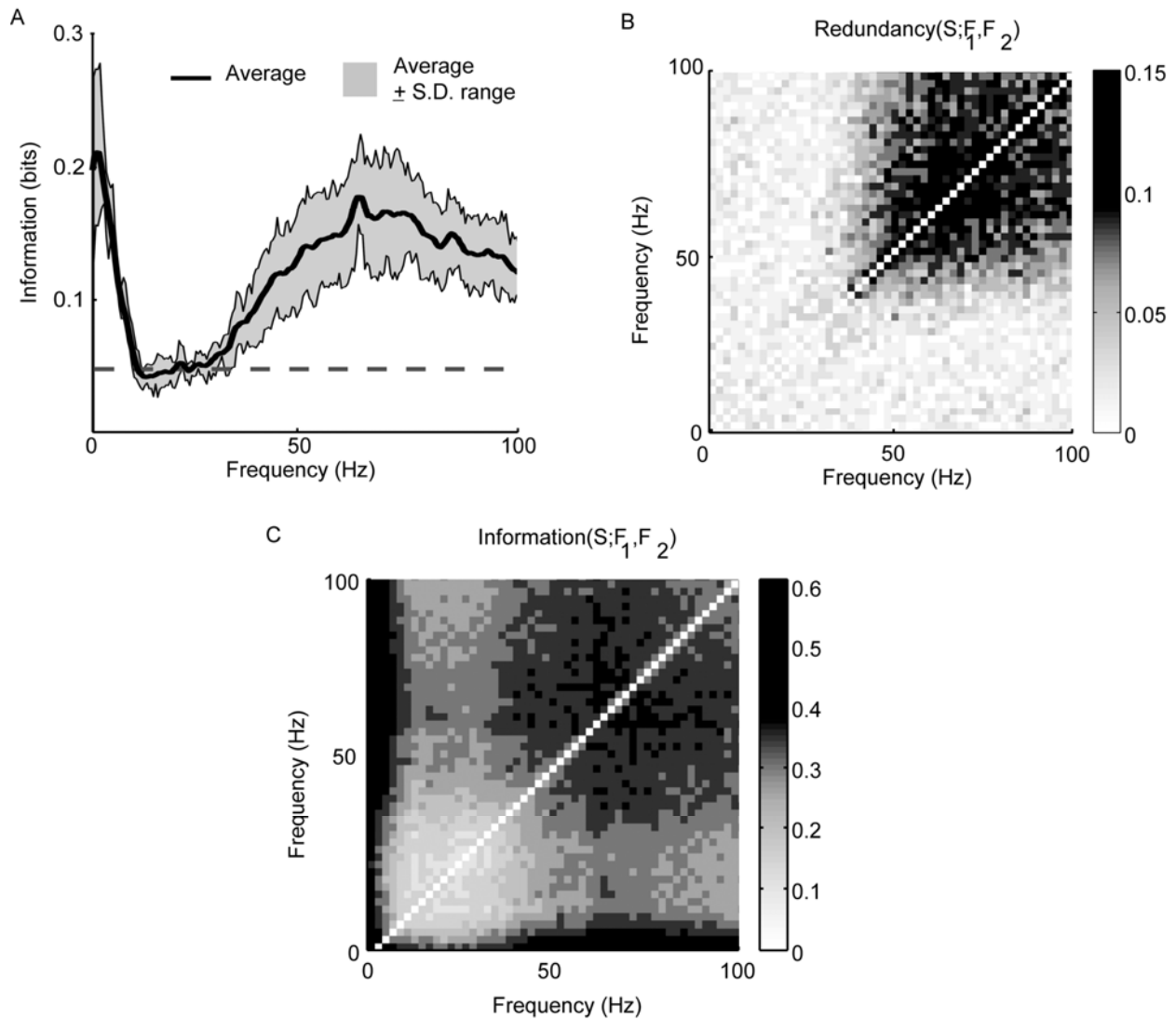

**Figure 3:** *Information about movie scenes carried by the power of LFPs at different frequencies.* (A) Information carried by different frequency bands (black line and gray area represent the mean ± STD area across recorded channels). The dashed line represents the p=0.05 (bootstrap test) significance line of information values. (B) Average across channels of the Redundancy of information carried by the power for all measured pairs of LFP frequencies. Note that frequencies in the gamma range are highly redundant. (B) Average across channels of the joint information carried by the LFP power at a given pair of frequency, plotted as function of the frequencies. The highest information is carried by the joint observation of the power of a low frequency and of a gamma frequency.

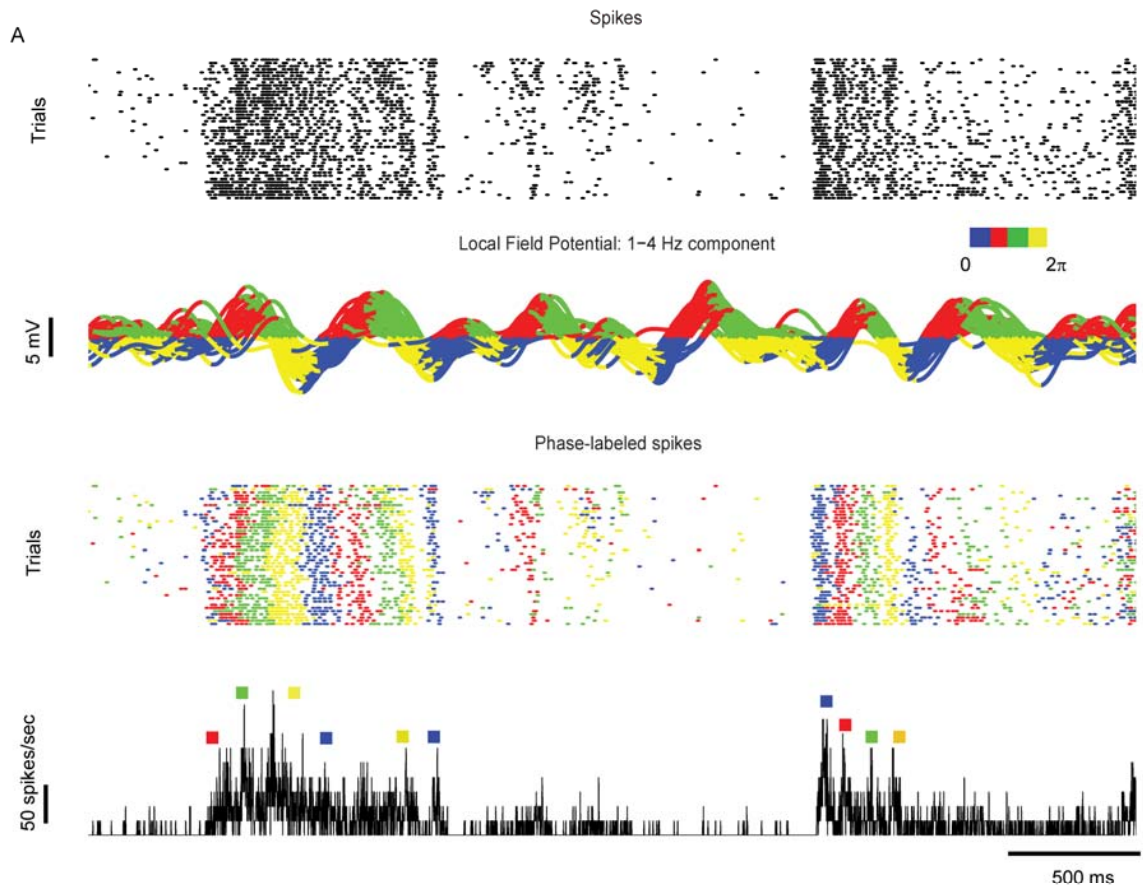

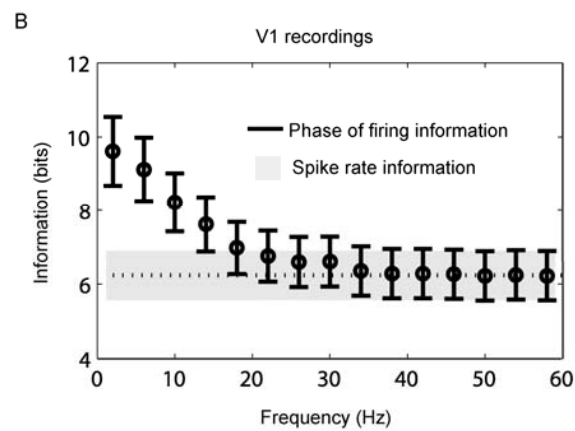

**Figure 4:** *Phase of firing coding.* (A) From top to bottom: raster plot of spike time, with each line representing a different trial and each dot representing the time of an action potential recorded from an example electrode; superimposition of the traces of delta band (1-4 Hz) of the LFP recorded from the same channel where the spikes were recorded, with the line color denoting the instantaneous LFP phase (discretized into phase quadrants as indicated in the inset); same spike times as displayed in top row, but with spike times colored according to the quadrant of the LFP phase at which they were emitted;

trial-average spike rate, with colored markers indicating the dominant LFP phase corresponding to different peaks in the trial-average time dependent firing rates. The fact that many peaks of the time dependent rate had similar amplitude but corresponded to spikes fired in different phase quadrants suggests that the phase of firing carries some additional information to that carried by spike rates alone. (B) Black circles show the information carried by the phase of firing code as function of the considered LFP frequency (mean±SEM over the data set). The black dashed line plots the mean over the data set of the spike rate information (SEM over data set indicated as gray area). Panel reproduced from (Montemurro et al., 2008) with permission.

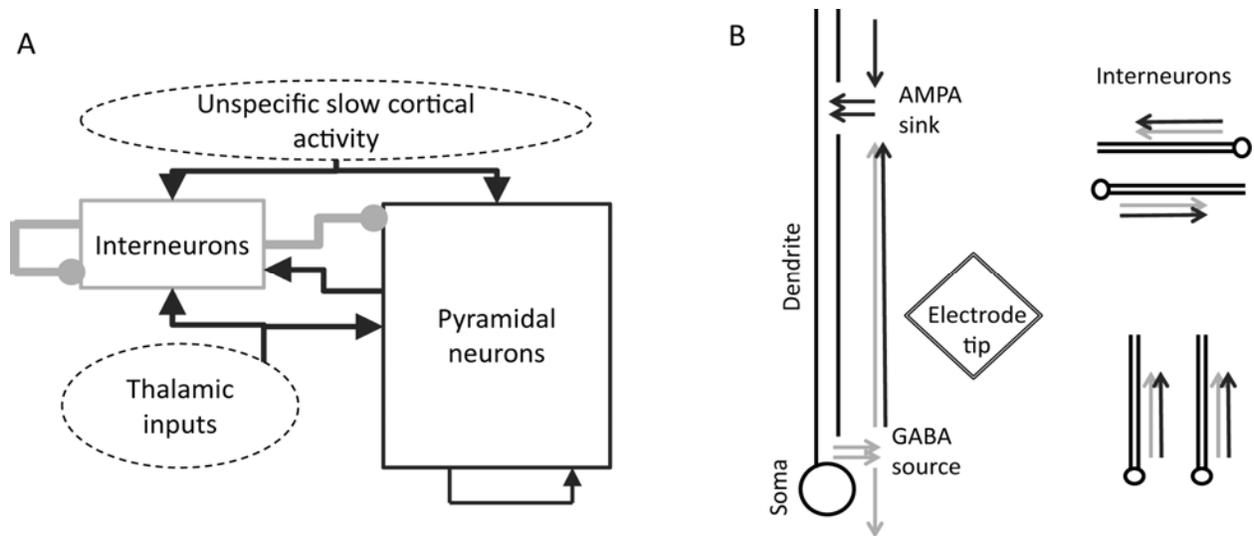

**Figure 5:** *Schematics of a randomly connected network of integrate-and-fire neurons.* (A) Schematics of network structure. The network was composed by 4000 AMPA neurons and 1000 GABA neurons. Directed pairs of neurons were connected with probability 0.2. The size of the arrows illustrates the strength of the different synaptic connections. In addition to recurrent interactions each neuron received two types of distinct external excitatory drives: a "thalamic" input carrying the simulated sensory information, and an "unspecific" input representing stimulus unrelated changes of ongoing activity and non-specific contributions from other areas. (B) Schematics of the computation of the simulated LFP. Left side: we computed the simulated LFP as the sum of the absolute values of AMPA and GABA currents because AMPA synapses are usually apical and GABA synapses are usually peri-somatic and thus their dipoles sum with the same sign along the dendrite. Right side: we summed only currents from synapses of pyramidal neurons because, due to their approximate open field arrangement, they neurons contribute more than inhibitory neurons, which have a much less regular dendritic spatial organization.

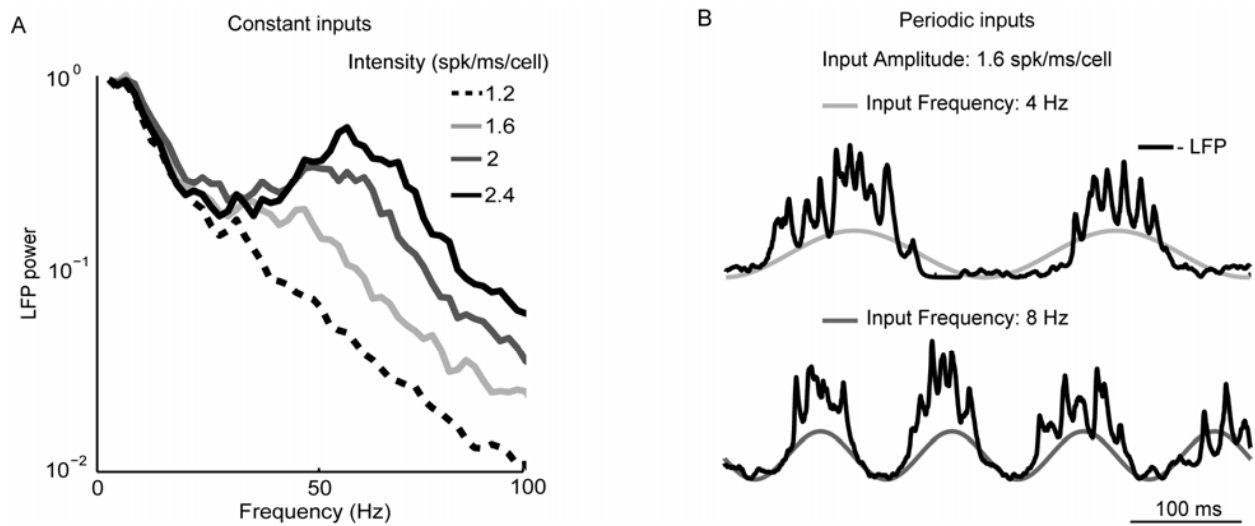

**Figure 6:** *Information processing in a recurrent network model injected with simple inputs.* (A) LFP spectral modulation when the network was injected with constant inputs of different intensity (see legend). The peak of the modulation occurred in the range [50-80 Hz] in the gamma band. (B) LFP dynamics when the network was injected with low-frequency periodic inputs (see legend). The low frequency component of the network LFP entrained to the input. The network also displayed cross frequency coupling between high frequency power and low frequency phase. Panels (A) adapted from (Mazzoni et al., 2008), panel (B) adapted from (Panzeri et al., 2010).

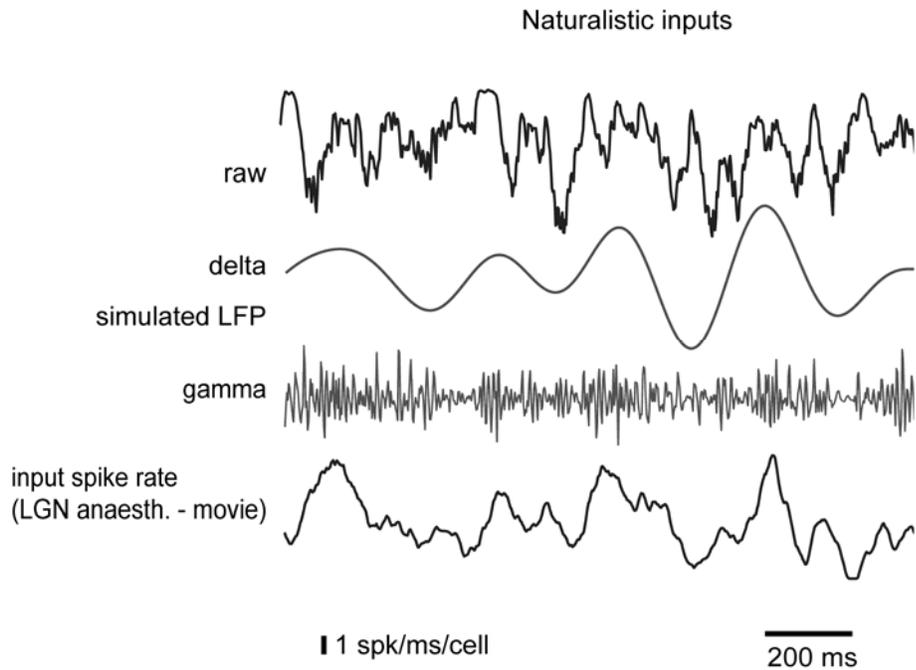

**Figure 7:** *Illustrations of the response of the simulated network to naturalistic input spike trains*. Network dynamics during naturalistic input: the delta [1-4 Hz] component of the LFP was entrained by the low-frequency structure of the input, while periods of large amplitude of the LFP gamma [30-100 Hz] power were associated to peaks in the input. Thus, even if naturalistic inputs displayed a broad range of timescales, low and gamma frequency bands were still associated to the same input features revealed in the study of constant and sinusoidal inputs. Panel adapted from (Mazzoni et al., 2010).

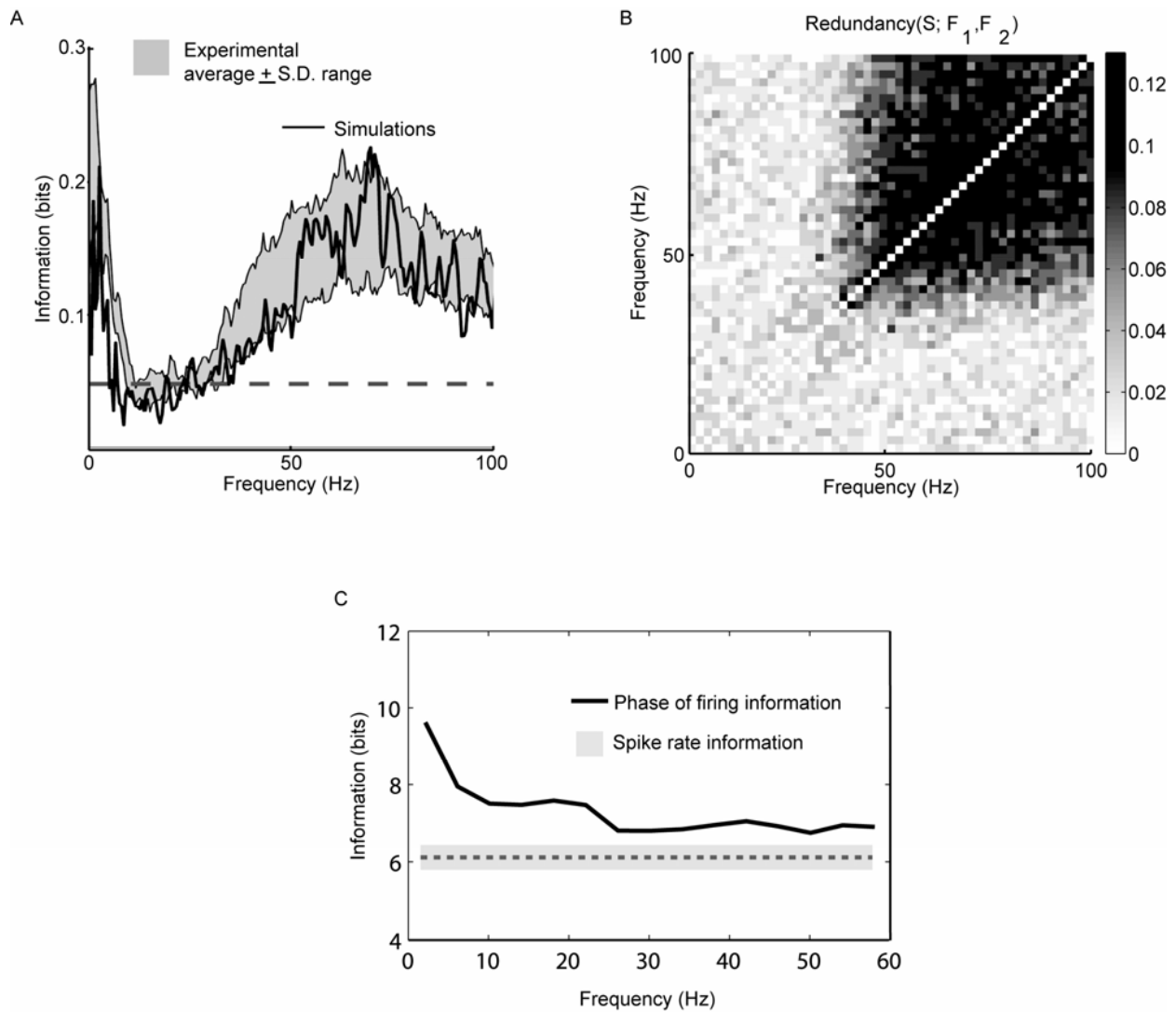

**Figure 8:** *The information carried by LFP power in a recurrent network model injected with naturalistic inputs*. (A) Information about different intervals of the dynamic input stimulus carried by the LFP power for both experimental recordings (gray area, representing the mean ± STD area across all channels) and simulations (black line). The dashed line represents the p=0.05 (bootstrap test) significance line of information values. (B) Redundancy of simulated LFP power for all frequency pairs. Results are consistent with experimental results: There were two peaks of information, one for low frequencies and one in the [50-100 Hz] frequency range in the gamma band; Low frequencies and gamma frequencies carried independent information; there was high redundancy within the gamma range. (C) Information carried by the summed activity of two simulated excitatory neurons from the network (total average rate: 5.6 spikes/sec) when naturalistic inputs were injected. Solid line represents the information carried by the phase of firing code, dashed line the information carried by the spike rate, and grey area the bootstrap significance (p<0.05) obtained with 40 permutations of the phase. Panels (A,B) adapted from (Mazzoni et al., 2008), and Panel C adapted from (Mazzoni et al., 2011).